\documentclass[aip,twocolumn,letterpaper]{revtex4}

\usepackage{fullpage}

\usepackage{graphics}
\usepackage{epsfig}

\begin{document}
\title{Particle acceleration and fast magnetic reconnection}
\author{Allen H. Boozer}
\affiliation{Columbia University, New York, NY  10027\\ ahb17@columbia.edu}

\begin{abstract}

Mathematics demonstrates that an ideally evolving magnetic field has an exponentially increasing sensitivity to non-ideal effects for all but truly exceptional evolutions.  On a time scale that depends only logarithmically on the magnitude of non-ideality, an evolving magnetic field will generally reach a state of fast magnetic reconnection.  The effects of fast magnetic reconnection proceed at a rate determined by Alfv\'enic, not resistive, physics.  The best known of these effects are associated with the transfer of magnetic field energy to the plasma and the conservation of magnetic helicity, which limits the energy transfer.  As will be shown, weak non-ideality implies helicity conservation in regions bounded by magnetic surfaces or rigid perfect conductors.  But, weak non-ideality makes the rapid transfer of energy subtle.  This transfer can be understood by the drive for Alfv\'en waves and two other effects, which are present even in an ideal evolution, an effective parallel electric field $\mathcal{E}_{||}$, which can accelerate particles despite the particle acceleration due to the true parallel electric field $E_{||}$ being negligible, and a coefficient $\nu_K$, which gives a rate for exponentiation of the kinetic energy of particle motion along the magnetic field.   Classical theories of reconnection are two-dimensional and exclude the exponentially increasing sensitivity that is robustly present for magnetic fields in three-dimensional space.

\end{abstract}

\date{\today} 
\maketitle


\section{Introduction}

By fast magnetic reconnection, we mean magnetic field lines changing their connections on a time scale determined by Alfv\'enic, not resistive, physics.  Fast magnetic reconnection is prevalent in both natural and laboratory plasmas \cite{Liu:2017}.

As shown by Newcomb\cite{Newcomb}, magnetic field lines cannot change their connections but move with a velocity $\vec{u}_\bot(\vec{x},t)$ if and only if the magnetic field obeys the ideal evolution equation \begin{equation}
\frac{\partial\vec{B}}{\partial t}=\vec{\nabla}\times (\vec{u}_\bot\times\vec{B}). \label{Ideal-ev}
\end{equation}  

The ideal evolution equation is deceptive because it has the mathematical property of containing the seeds of its own destruction when the magnetic field depends on three, though not on two, spatial coordinates \cite{Boozer:ideal-ev}.   What is meant is that the connections-breaking magnetic field,  $\vec{B}_{ni}$, is proportional to the deviation $\mathcal{E}_{ni}$ of the electric field from the ideal form multiplied by a factor $\Lambda_u$ that increases exponentially in time; $\ln{\Lambda_u}/t \equiv \lambda_u$ as $t\rightarrow\infty$.  The average rate at which streamlines of $\vec{u}_\bot$ e-fold apart is the Lyapunov exponent, $\lambda_u$, of the magnetic field line velocity.  The Lyapunov exponent of a generic flow is of order the largest element in the three-by-three matrix $\vec{\nabla}\vec{u}_\bot$, so $\lambda_u\sim u_\bot/a$, where $a$ is a distance scale for variations in the flow.  

The two-dimensional case, which was considered by Longcope and Strauss \cite{Longcope-Strauss}, is only consistent with an exponential increase in the sensitivity when the forces exerted by the plasma on the magnetic field also change exponentially.  Appendix \ref{sec:L-coord} and  \cite{Boozer:ideal-ev} show an exponentiation of forces is not required in three dimensions because of the presence of a middle singular value $\Lambda_m$ of the Singular Value Decomposition of the Jacobian matrix of Lagrangian coordinates. 

  As shown in Section \ref{sec:E}, away from null-points of the magnetic field, the electric field has a general form $\vec{E}+\vec{u}_\bot\times\vec{B}=-\vec{\nabla}\Phi +\mathcal{E}_{ni}\vec{\nabla}\ell$ with $\ell$ the distance along a magnetic field line.  When this form is used in Faraday's law, Equation (\ref{Ideal-ev}) is obtained when the deviation from ideality $\mathcal{E}_{ni}$ is zero.

  The time required for an evolving magnetic field to reach a state of fast magnetic reconnection, which is called the trigger time, is
\begin{eqnarray}
\tau_{trig} &\sim&  \frac{a}{u_\bot} \ln\left(\frac{u_\bot B}{ \mathcal{E}_{ni}}\right) \label{tau_trig}
\end{eqnarray}
When resistivity is the dominant cause of non-ideality, $\mathcal{E}_{ni}\approx \eta j_{||}\approx (\eta/\mu_0)B/a$, and
\begin{eqnarray}
\frac{u_\bot B}{ \mathcal{E}_{ni}} \approx \frac{\tau_\eta}{a/u_\bot}\equiv R_m,
\end{eqnarray}
where $\tau_\eta \equiv a^2/(\eta/\mu_0)$ is the resistive time scale.  $R_m$ is the magnetic Reynolds number $R_m$ of Zweibel and Yamada \cite{Zweibel:review} or the ideality $\Im$ of \cite{Boozer:ideal-ev}.   The values that Zweibel and Yamada gave for $R_m$ were $10^4$ to $10^8$ in large laboratory plasmas, $10^8$ to $10^{14}$ in the Sun, and $10^{15}$ to $10^{21}$ in the  interstellar medium of galaxies.

\begin{itemize}
\item  No matter how small the deviation $\mathcal{E}_{ni}$ from ideality may be, a magnetic field can change its connections on a time scale that increases only logarithmically as $\mathcal{E}_{ni}$ goes to zero.
\end{itemize}

A large difference in the expected current density separates traditional reconnection theory, which excludes exponentiation, from the expectation when exponentiation is included.  Traditional theories of magnetic reconnection have been two dimensional and relied on current sheets: from the classic work of Sweet \cite{Sweet:1958} and Parker \cite{Parker:1957} in the 1950's to work covered in recent reviews \cite{Zweibel:review,Loureiro:2016}.   

Schindler, Hesse, and Birn  \cite{Schindler:1988} gave the classical view of magnetic reconnection in three dimensions.  They assumed as do all traditional theories \cite{Liu:2017} that the part of the magnetic flux that is reconnecting, $\psi_p$, must be dissipated by the non-ideal electric field, $\partial \psi_p/\partial t=\mathcal{E}_{ni}L$, where $L$ is the length of the system.  This is not required in three dimensions in the presence of a guide field;  the reconnecting flux is mixed, not destroyed.   The discussion is simpler in  toroidal geometry, as in a tokamak, where the reconnecting flux can be identified with the poloidal magnetic flux $\psi_p$.  In an ideal evolution \cite{Boozer:RMP}, the poloidal $\psi_p$ and the toroidal $\psi_t$ magnetic fluxes are tied together $\partial \psi_p(\psi_t,t)/\partial t=0$.  But as shown in Section \ref{sec:A.B cons}, a fast magnetic reconnection conserves helicity, within the reconnecting region, or  equivalently $\int \psi_p(\psi_t,\theta,\varphi) d\psi_td\theta d\varphi$ where $\theta$ and $\varphi$ are the two angles of a torus  \cite{Boozer:e-runaway2019}.   The poloidal flux is the Hamiltonian for magnetic field lines, $d\psi_t/d\varphi = -\partial\psi_p/\partial \theta$ and $d\theta/d\varphi = \partial\psi_p/\partial \psi_t$.  Before a fast magnetic reconnection, each magnetic field line lay in a constant-$\psi_p$ magnetic surface.  After the reconnection a large region of stochastic field lines is formed, which can lie between inner to an outer magnetic surface with fluxes $\psi_p^{(i)}$ and $\psi_p^{(o)}$.  A typical field line in the stochastic region samples all values of $\psi_p$ with $\psi_p^{(i)}<\psi_p<\psi_p^{(o)}$.

Fast magnetic reconnection is a quasi-ideal process conserving magnetic helicity, which constrains the fraction of the magnetic field energy that can be released. During the fast magnetic reconnection that frequently occurs during tokamak disruptions, helicity conservation severely limits the change in the magnetic energy  \cite{Boozer:e-runaway2019}.  The important effect is the flattening of the current profile, $j_{||}/B$, on the time scale for shear Alfv\'en waves to spread along the magnetic field lines.

 Although current sheets generically form in a near-ideal evolution, the current density increases exponentially slower than does the non-ideal part of the magnetic field \cite{Boozer:ideal-ev}, and is subdominate to the $e^{\lambda_u t}$ exponentiation as a cause for reconnection when $R_m\rightarrow\infty$.  The current sheets that form reach a characteristic current density $j = (B/\mu_0 a)/\ln(R_m)$ when fast reconnection occurs.  In linear theory, the current sheets are extended along the magnetic field and ribbon-like across, exponentially broad in one direction and exponentially narrow in the other \cite{Boozer:ideal-ev}.  These exponentially broad and narrow directions have an extremely complicated form in Cartesian coordinates.  The characteristic spatial scale of regions with strong currents is the harmonic mean of the two, $a/\ln(R_m)$.  As discussed in the Summary, Section \ref{Summary}, this is the characteristic scale of currents resolved in recent solar observations of coronal mass ejections \cite{Gou:2019}.  
     
General or generic behavior means that even in special cases in which it does not occur a small perturbation can restore that behavior.  The features of the near-ideal three-dimensional evolution of magnetic fields are generic.  For example, continuous spatial symmetries can make the Lyapunov exponent of the flow $\gamma_u=0$, but then small perturbations can make $\gamma_u>0$.  

The generic behavior of the near-ideal evolution of magnetic fields in three dimensions provides answers to the four questions that are required for a practical understanding of fast magnetic reconnection phenomena:

\begin{enumerate}

\item Why does the near-ideal evolution of natural and laboratory magnetic fields robustly lead to states of fast magnetic reconnection independent of the drive and of the initial state?  

\item  What is the characteristic time required to reach a state of fast magnetic reconnection?

\item What is the explanation of the effects produced by fast magnetic reconnection, which are primarily associated with magnetic helicity conservation and an energy transfer from the magnetic field to the plasma?  

\item  Why does the Alfv\'en speed define the time scale during which the effects produced by fast magnetic reconnection occur?

\end{enumerate}

Helicity conservation is generally far less limiting in the energy loss from the magnetic configurations that arise in space and astrophysical plasmas than in tokamaks.  Nevertheless, the energy transfer in a fast magnetic reconnection occurs under the constraint of a quasi-ideal process, $\mathcal{E}_{ni}\rightarrow0.$  Under this constraint, a large energy transfer from the magnetic field to the plasma can occur through (a) the production and damping of shear Alfv\'en waves, Section \ref{sec:A.B cons} and Appendix \ref{sec:Alfven}, and (b) the interaction between the guiding center drift velocity and the perpendicular electric field.   This interaction can increase the parallel kinetic energy $K_{||}=\frac{1}{2}mv_{||}^2$ of individual particles in two ways.  (1) A coefficient $\nu_K$, Equation (\ref{nu_K def}), can exponentiate $K_{||}$.  (2) An effective  parallel electric field $\mathcal{E}_{||}$, Equation (\ref{eff.E-field}), can accelerate particles in a way that is analogous to a true parallel electric field $E_{||}$ even when $\mathcal{E}_{ni}\rightarrow0$.  The energization associated with $\nu_K$ appears to be the more important.  Both $\nu_K$ and $\mathcal{E}_{||}$ are non-zero in an ideal evolution, $\mathcal{E}_{ni}=0$, but their magnitudes depend on the rate at which the magnetic field is evolving, and that rate is typically much larger during a period in which fast magnetic reconnection is occurring.

This paper is not the first to point out that particle drifts can interact with the electric field perpendicular to the magnetic field to increase the kinetic energy in the parallel motion  of charged particles.  The magnitude of $\vec{E}_\bot$ depends on the magnetic field line velocity $\vec{u}_\bot$ and can be large even in an ideal evolution.  Browning and Vekstein \cite{Browning:2001} carried out extensive calculations of this effect. Dahlin, Drake, and Swisdak \cite{Drake:2017} have found that the energy transfer though $E_{||}$ need not be dominant even when $E_{||}$ is large.  In their analysis, the important transfer was an exponentiation in $K_{||}$ by a term proportional to $\vec{u}_E\cdot\vec{\kappa}$, where $\vec{u}_E\equiv \vec{E}\times\vec{B}/B^2$.  As shown in Section \ref{sec:dK_||/dt}, the origin of $\nu_K$ stems from the term $\vec{u}_\bot\cdot\vec{\kappa}$ in the energy transfer.  

Section \ref{sec:E} derives a general expression for the electric field in spatial regions without magnetic nulls and the rate at which energy is removed from the magnetic field.  Section \ref{KE ev} gives Northrop's formula for the rate of change in the kinetic energy of a particle, $K$, in the guiding-center-drift approximation, and the expression for $dK/dt$ that results from the general expression for the electric field.   Section \ref{sec:dK_||/dt} derives the rate of change the kinetic energy in particle motion along the magnetic field lines, $K_{||}$.  The perpendicular kinetic energy $K_\bot=\mu B + m u_\bot^2/2$, with $\mu$ the magnetic moment is not subject to the same type of change.  Section \ref{sec:A.B cons} addresses the conservation of magnetic helicity.  Section \ref{Summary} summarizes the paper.  Appendix \ref{sec:L-coord} discusses Lagrangian coordinates.  Appendix \ref{sec:dK/dt} gives algebraic details of the derivation of the $dK/dt$ equation of Section \ref{KE ev}.  Appendix \ref{u.kappa} gives algebraic details for the derivation of the formula for the dot product of the magnetic field line velocity and the field line curvature, which is used in Section \ref{sec:dK_||/dt}.  Appendix \ref{sec:Alfven} derives the equation for Alfv\'en waves driven by a non-zero $\vec{B}\cdot\vec{\nabla}(j_{||}/B)$.  Appendix \ref{corona} discusses runaway electrons, which may explain the solar corona.   Runaway electrons are expected in plasmoid models during fast magnetic reconnections in parameter regimes similar to those of the solar corona but can arise more generally.


\section{The electric field and the magnetic energy transfer \label{sec:E}}

The part of the electric field that prevents the magnetic field evolution from being ideal is a constant $\mathcal{E}_{ni}$ along any field line that does not intercept a magnetic field null \cite{Boozer:e-runaway2019}.

A general Ohm's law has the form $\vec{E}+\vec{v}\times\vec{B} =\vec{\mathcal{R}}$, where $\vec{v}$ is the plasma velocity \cite{Schindler:1988}.  In non-relativistic theory, which is used in this paper, $\vec{\mathcal{R}}$ is the electric field in a frame that moves with the plasma.   Let $\Phi$ be a solution to $\vec{B}\cdot\vec{\nabla}\Phi=-\vec{B}\cdot\vec{\mathcal{R}}-\mathcal{E}_{ni} B$, where $\mathcal{E}_{ni}$ is constant along each magnetic field line.  Define a velocity perpendicular to the magnetic field $\vec{u}_\bot$ by
\begin{eqnarray}
&&\vec{v}_\bot =  \vec{u}_\bot +\vec{B}\times \frac{\vec{\mathcal{R}}+\vec{\nabla}\Phi}{B^2}, \hspace{0.2in}\mbox{then}\hspace{0.2in}\\
&&\vec{E}+\vec{u}_\bot\times\vec{B}=-\vec{\nabla}\Phi +\mathcal{E}_{ni}\vec{\nabla}\ell. \label{E}
\end{eqnarray}
The distance along a magnetic field line is $\ell$, and
\begin{equation}
\mathcal{E}_{ni} \equiv \frac{\int \vec{E}\cdot\vec{B}d\ell}{\int d\ell},
\end{equation}
where both integrals are calculated as the limits of integration go to plus/minus infinity or from one contact with a perfectly conducting wall to another.

Poynting's theorem, Equation (\ref{Poynting eq}), implies  $\vec{j}\cdot\vec{E}$ is the rate per unit volume at which energy is removed from the magnetic field.    Using Equation (\ref{E}) for the electric field,
\begin{equation}
\vec{j}\cdot\vec{E} = \vec{u_\bot}\cdot\vec{f}_L + \mathcal{E}_{ni}\vec{j}\cdot\vec{\nabla}\ell,
\end{equation}
where $\vec{f}_L=\vec{j}\times\vec{B}$ is the Lorentz force.  The first term on the righthand side of the $\vec{j}\cdot\vec{E}$ equation gives the ideal transfer and the second the dissipative loss of energy by the magnetic field.  All of the energy lost by the magnetic field is eventually transferred to the plasma, but the ideal transfer includes the transfer of energy to Alfv\'en waves, which are then damped by the plasma as discussed in Section \ref{sec:A.B cons}.


\section{Energy transfer to charged particles \label{KE ev} }

The ideal energy transfer from the magnetic field includes the direct transfer to the kinetic energy of charged particles.  When the gyroradius of a non-relativistic changed particle is small, its kinetic energy is
\begin{equation}
K = \frac{m}{2} v_{||}^2 +\mu B + \frac{m}{2} u_{\bot}^2, \label{K}
\end{equation}
where $\mu$ is the adiabatically conserved magnetic moment.  In his paper on the guiding center velocity, Northrop \cite{Northrop:1963} found the time derivative of the kinetic energy is
\begin{equation}
\frac{dK}{dt} = q\vec{v}_g\cdot\vec{E} + \mu \left(\frac{\partial B}{\partial t}\right)_{\vec{x}}, \label{Northrop dK/dt}
\end{equation}
where $\vec{v}_g$ is the velocity of the guiding center of the particle.  Northrop expression for $\vec{v}_g$, Equation (\ref{v_g Northrop}), includes terms that are not retained in the standard textbook expressions.  These terms are retained in the derivation of $dK/dt$ given in Appendix \ref{sec:dK/dt} since they can be important;
\begin{eqnarray} 
&&\frac{dK}{dt} = \frac{m}{2} \frac{d u_\bot^2}{d t} + \mu \left(\frac{\partial B}{\partial t}\right)_L - qv_{||}\hat{b}\cdot\vec{\nabla}\Phi  \nonumber\\ 
&& +  mv_{||}^2\vec{u} _\bot\cdot\vec{\kappa}  + v_{||} m\vec{u}_\bot \cdot \left(\frac{\partial\hat{b}}{\partial t}\right)_L +q v_{||}\mathcal{E}_{ni}.  \hspace{0.1in} \label{dK/dt eq}
\end{eqnarray}

 Three types of time derivatives must be distinguished.  The time derivative $(\partial f/\partial t)_{\vec{x}}$ is taken at a fixed spatial point.  The Lagrangian or convective derivative, 
 \begin{equation}
 \left(\frac{\partial f}{\partial t} \right)_L \equiv \left(\frac{\partial f}{\partial t} \right)_{\vec{x}} + \vec{u}_\bot\cdot\vec{\nabla}f, \label{Lagrangian derivative}
 \end{equation} 
 is taken in a frame moving with the magnetic field lines.  The total time derivative is taken in the frame of the charged particle, 
 \begin{equation}
 \frac{df}{dt} \equiv  \left(\frac{\partial f}{\partial t} \right)_{\vec{x}} + (v_{||}\hat{b}+ \vec{u}_\bot)\cdot\vec{\nabla}f,
 \end{equation}
 ignoring terms that go to zero as $qB/m$ goes to infinity.  Keeping those terms would require second order drifts for consistency.  The velocity of the particle along the magnetic field $v_{||}\hat{b}$ and the velocity with which it is carried by the motion of the magnetic field lines $\vec{u}_\bot$ are the only two components of the velocity that do not go to zero as the gyrofrequency $qB/m\rightarrow\infty$ with everything else held fixed.


\section{Evolution of the parallel kinetic energy \label{sec:dK_||/dt} }

The kinetic energy of a particle can be separated into the part given by the velocity of the particle perpendicular to the magnetic field $K_\bot=\mu B + \frac{m}{2}u_\bot^2$ and the part $K_{||}=\frac{m}{2} v_{||}^2$ associated with its velocity along the magnetic field with $K=K_{||}+K_\bot$.  Here the focus is on the evolution of the parallel kinetic energy, for $K_{\bot}$ is already expressed in easily determined quantities, $B$ and $u_\bot^2$.

The kinetic energy, Equation (\ref{K}), can be written $K=K_{||} + \mu B + \frac{m}{2}u_\bot^2$.  The parallel kinetic energy is
\begin{eqnarray}
\frac{dK_{||}}{dt} &=& \frac{dK}{dt} -\mu \frac{dB}{dt} - \frac{m}{2} \frac{d u_\bot^2}{dt}.
\end{eqnarray}

The difference between $dB/dt$ and $(\partial B/\partial t)_L$ is $dB/dt -(\partial B/\partial t)_L = v_{||} \hat{b}\cdot \vec{\nabla}B$, so
\begin{eqnarray}
\frac{dK_{||}}{dt}&=& -\mu v_{||}  \hat{b}\cdot\vec{\nabla}B + q v_{||} ( \mathcal{E}_{ni}  -\hat{b}\cdot\vec{\nabla}\Phi)  \nonumber \\
&& + v_{||} m\vec{u}_\bot \cdot \left(\frac{\partial\hat{b}}{\partial t}\right)_L  + mv_{||}^2\vec{u}_\bot\cdot\vec{\kappa}. \label{K_|| ev} \hspace{0.2in}
\end{eqnarray}

The last two terms in Equation (\ref{K_|| ev}) can be placed in a more useful form through further analysis.  An expression for $\vec{u}_\bot\cdot\vec{\kappa}$ is obtained in Appendix \ref{u.kappa}.  The analysis of the $\vec{u}_\bot \cdot(\partial\hat{b}/\partial t)_L$ term and further analysis of the $\vec{u}_\bot\cdot\vec{\kappa}$ term will require constraints that become clear with the use of Lagrangian coordinates, Appendix \ref{sec:L-coord}.



\subsection{Effective parallel electric field, $\mathcal{E}_{||}$ \label{Eff. E_||} }

Equation (\ref{K_|| ev}) for the time derivative of the parallel kinetic energy and Equation (\ref{u-kappa}) for $\vec{u}_\bot\cdot\vec{\kappa}$ can be combined to obtain a general expression for the time derivative of the parallel kinetic energy in the small-gyroradius limit,
\begin{eqnarray}
\frac{dK_{||}}{dt} &+& 2K_{||}\left(\left(\frac{\partial \ln(\Lambda_m)}{\partial t} \right)_L+ \frac{ \vec{B}\cdot( \vec{\nabla}\ell\times\vec{\nabla}\mathcal{E}_{ni}) }{B^2/\mu_0}\right) \nonumber\\
&& = -\mu v_{||}  \hat{b}\cdot\vec{\nabla}B + q v_{||} ( \mathcal{E}_{ni}  -\hat{b}\cdot\vec{\nabla}\Phi)  \nonumber \\ 
&& \hspace{0.2in} + v_{||} m\vec{u}_\bot \cdot \left(\frac{\partial\hat{b}}{\partial t}\right)_L \nonumber  \\
&&\equiv   qv_{||}\mathcal{E}_{||}.  \label{dK_||/dt equation}
\end{eqnarray}
The effective parallel electric field is
\begin{eqnarray}
\mathcal{E}_{||} &\equiv& \mathcal{E}_{ni} -\hat{b}\cdot\vec{\nabla}\left( \frac{\mu}{q}B+\Phi\right)  +m\vec{u}_\bot \cdot \left(\frac{\partial\hat{b}}{\partial t}\right)_L. \hspace{0.3in}  \label{mathcal(E) deff}
\end{eqnarray}

Equation (\ref{mathcal(E) deff}) for the effective parallel electric field can be simplified.  The unit vector along the magnetic field can be written as $\hat{b}=\hat{b}_I+\vec{b}_{ni}$, where $\vec{b}_{ni}$ is the assumed small change in the direction of the magnetic field produced by non-ideal effects and $\hat{b}_I$ is the unit vector along the ideal part of the magnetic field.  Appendix \ref{sec:L-coord} shows $\hat{b}_I=\hat{M}$ in Lagrangian coordinates.
Equation (\ref{u.dbI/dt}) implies
\begin{equation}
m\vec{u}_\bot \cdot \left(\frac{\partial\hat{b}}{\partial t}\right)_L=\hat{b}_I\cdot \vec{\nabla}\frac{mu_\bot^2}{2} + m\vec{u}_\bot \cdot \left(\frac{\partial\vec{b}_{ni} }{\partial t}\right)_L.
\end{equation}
The other $\hat{b}$'s in Equation (\ref{mathcal(E) deff}) can be replaced by $\hat{b}_I$.  The effective parallel electric field is then
\begin{eqnarray}
&& \mathcal{E}_{||}=\mathcal{E}_{ni} -\hat{b}_I\cdot\vec{\nabla}\Phi_{eff} +\frac{m}{q} \vec{u}_\bot\cdot  \left( \frac{\partial\vec{b}_{ni}}{\partial t}\right)_L; \hspace{0.2in} \label{eff.E-field} \\
&& \Phi_{eff} \equiv \Phi +\frac{\mu B_I}{q} - \frac{mu_\bot^2}{2q}.
\end{eqnarray}


\subsection{Equation for $dK_{||}/dt$ }

The complicated term that multiplies $K_{||}$ in the first line of Equation (\ref{dK_||/dt equation}) can also be simplified.  
\begin{eqnarray}
\frac{\vec{B}\cdot ( \vec{\nabla}\ell\times \vec{\nabla}\mathcal{E}_{ni})}{B^2}& = &\ \frac{\partial \ell}{\partial\alpha_I} \frac{\partial \mathcal{E}_{ni}}{\partial\beta_I} -  \frac{\partial \ell}{\partial\beta_I} \frac{\partial \mathcal{E}_{ni}}{\partial\alpha_I} \hspace{0.1in} \label{ell-E_|| term}
\end{eqnarray}
where $\alpha_I$ and $\beta_I$ are the Clebsch potentials of the ideal magnetic field, $\vec{B}_I=\vec{\nabla}\alpha_I\times\vec{\nabla}\beta_I$.  As discussed in \cite{Boozer:ideal-ev}, $\alpha_I$ and $\beta_I$ can be locally chosen so the derivatives with respect to $\alpha_I$ are exponentially large and derivatives with respect to $\beta_I$ are exponentially small, and their combination has a weak dependence on time.  Consequently, this term is rarely important.

The time derivative of the parallel energy can then be written as 
\begin{eqnarray}
\frac{dK_{||}}{dt} & = & -\left( \frac{\partial\ln\Lambda_m^2}{\partial t} \right)_L K_{||} +qv_{||}\mathcal{E}_{||}, \mbox{    and   } \hspace{0.2in} \\
 \frac{d\ln\Lambda_m^2}{d t} &=& \left( \frac{\partial\ln\Lambda_m^2}{\partial t} \right)_L + v_{||}\hat{b}_I\cdot\vec{\nabla}\ln\Lambda_m^2. \hspace{0.2in}
\end{eqnarray}
Consequently, the rate of change of the parallel kinetic energy is 
\begin{eqnarray}
&& \frac{dK_{||}}{dt} +\frac{d\ln\Lambda_m^2}{d t} K_{||} = \nu_K K_{||} + qv_{||}\mathcal{E}_{||},   \hspace{0.2in} \label{dK||/dt eq}\\
\end{eqnarray} 
where the growth rate of $K_{||}$ can be written as
\begin{eqnarray}
\nu_K &=& v_{||}\hat{b}_I\cdot\vec{\nabla}\ln\Lambda_m^2. \label{nu_K def}
\end{eqnarray} 
When $\mathcal{E}_{||}=0$, 
\begin{equation}
\frac{d\ln(\Lambda_m^2K_{||})}{dt} =v_{||}\hat{b}_I\cdot\vec{\nabla}\ln\Lambda_m^2.
\end{equation} 

The large and longterm changes in $K_{||}$ are given by the time averaged product of variations in $v_{||}$ and variations in $\hat{b}_I\cdot\vec{\nabla}\ln\Lambda_m^2$.  Only variations can give a systematic acceleration since $(\int \hat{b}_I\cdot\vec{\nabla}\ln\Lambda_m^2 d\ell)/(\int d\ell)$ goes to zero as the range of integration goes to infinity.

When both $\nu_K$ and $\mathcal{E}_{||}$ are zero, $\Lambda_m^2 K_{||}$ is independent of time.  A heuristic argument relates this to the conservation of longitudinal action, $\oint v_{||}d\ell$.  Ignoring the non-ideal term, Equation (\ref{u-kappa}) is $\vec{u}_\bot\cdot\vec{\kappa} =-(\partial \ln(\Lambda_m)/\partial t )_L$.  Let $r$ be the radius of curvature of a magnetic field line, then the curvature $\vec{\kappa} = -\hat{r}/r$ and $\vec{u}_\bot\cdot\vec{\kappa}=-(\partial r/\partial t)_L/r$.  Consequently, $\Lambda_m\propto r$ and $\Lambda_m^2 K_{||}\propto r^2 v_{||}^2$ while  $\oint v_{||}d\ell \sim v_{||} r$.  A related argument is given in Dahlin, Drake, and Swisdak \cite{Drake:2017}.

The creation of high-energy tails on particle distribution functions is not a proof that reconnection has taken place; both $\nu_K$ and $\mathcal{E}_{||}$ can be non-zero when a magnetic field is evolving ideally.  An ideal evolution can lead to magnetic field configurations that are subject to ideal instabilities, which can cause the magnetic fields to evolve on Alfv\'enic time scale despite a perfect conservation of magnetic field line connections.  Nevertheless, the preservation of magnetic field line connections is a far stronger constraint than is helicity conservation, and the fractional change in the magnetic energy is generally much smaller in an ideal instability than in a fast magnetic reconnection.


\section{Magnetic helicity conservation \label{sec:A.B cons} }

The concept of magnetic helicity was introduced in 1958 by Woltjer \cite{Woltjer:1958} and became a well-known concept after it was used 1974 by Taylor to explain the reversal of the toroidal magnetic field during turbulent relaxations of the reversed field pinch \cite{Taylor:1974}.

As $\mathcal{E}_{ni}\rightarrow0$, the loss of  magnetic helicity $\int\vec{A}\cdot\vec{B}d^3x$ in a volume defined by magnetic surfaces or by rigid perfect conductors goes to zero during the time to trigger a fast magnetic reconnection, $\tau_{trig}$, Equation (\ref{tau_trig}).   During that time, the fractional change in the helicity is of order $(\ln R_m)/R_m$, where $R_m=(u_\bot B)/\mathcal{E}_{ni}$.  This is a different statement on helicity loss than the bound obtained by Berger \cite{Berger:1984}.   He found, when properly normalized, that the dissipation of magnetic helicity in a turbulent plasma is smaller than the dissipation of magnetic energy.

A calculation of the helicity change begins with the time derivative of $\vec{A}\cdot\vec{B}$.  Using $\vec{E}=-\partial \vec{A}/\partial t - \vec{\nabla}\Phi$ with $\vec{B}=\vec{\nabla}\times\vec{A}$,
\begin{eqnarray}
&&\left(\frac{\partial \vec{A}\cdot\vec{B}}{\partial t}\right)_{\vec{x}}=-(\vec{E}+\vec{\nabla}\Phi)\cdot\vec{B}-\vec{A}\cdot\vec{\nabla}\times\vec{E} \hspace{0.2in}\\
&&\hspace{0.2in} = -2\vec{E}\cdot\vec{B}-\vec{\nabla}\cdot(\Phi \vec{B}) +\vec{\nabla}\cdot(\vec{A}\times\vec{E}).
\end{eqnarray}
Using Equation (\ref{E}) for the electric field
\begin{eqnarray}
\vec{A}\times\vec{E}&=&-(\vec{A}\cdot\vec{B})\vec{u}_\bot +(\vec{A}\cdot\vec{u}_\bot)\vec{B}+\vec{\nabla}\times(\Phi\vec{A})\nonumber\\
&&-\Phi\vec{B}-(\mathcal{E}_{ni}\vec{\nabla}\ell)\times\vec{A}
\end{eqnarray}
Consequently,
\begin{eqnarray}
&&\left(\frac{\partial \vec{A}\cdot\vec{B}}{\partial t}\right)_{\vec{x}}=- 2(\vec{E}+\vec{\nabla}\Phi)\cdot\vec{B} -\vec{\nabla}\cdot\vec{\mathcal{F}} \hspace{0.1in}\mbox{   where   } \hspace{0.2in}\\ \nonumber\\
&&\vec{\mathcal{F}}= (\vec{A}\cdot\vec{B})\vec{u}_\bot-(\vec{A}\cdot\vec{u}_\bot)\vec{B} +(\mathcal{E}_{ni}\vec{\nabla}\ell)\times\vec{A}.\hspace{0.3in}
\end{eqnarray}
The evolution of the magnetic helicity in a volume is
\begin{eqnarray}
\left(\frac{\partial \int\vec{A}\cdot\vec{B}d^3x}{\partial t}\right)_{\vec{x}}=-\int\mathcal{E}_{ni}Bd^3x -\oint \vec{\mathcal{F}}\cdot d\vec{a}. \hspace{0.1in}
\end{eqnarray}
The volumetric term $\int\mathcal{E}_{ni}Bd^3x$ gives the stated fractional change in helicity, $\sim (\ln R_m)/R_m$, during the time required to trigger a fast magnetic reconnection.

When the bounding surfaces are magnetic surfaces, a surface on which the normal component of $\vec{B}$ vanishes, the only surface term that does not go to zero as $\mathcal{E}_{ni}$ does is $\oint (\vec{A}\cdot\vec{B})\vec{u}_\bot \cdot d\vec{a} $.  This term implies the magnetic helicity moves with the magnetic field lines, hardly an unexpected result.   The surface term $\oint (\mathcal{E}_{ni}\vec{\nabla}\ell)\times\vec{A}\cdot d\vec{a}$ can be shown to be the helicity input by the surface loop voltage--the time derivative of the poloidal magnetic flux with the toroidal magnetic flux held constant.

When the bounding surface is a rigid perfect conductor, $\vec{u}_\bot=0$, then $\oint \vec{\mathcal{F}}\cdot d\vec{a}=0$.  The component of the magnetic field normal to a perfect conductor need not be zero, but the velocity of the magnetic field lines in a rigid perfect conductor must be zero.

Static force balance is broken when fast magnetic reconnection joins field lines that have different values of $j_{||}/B\equiv\vec{j}\cdot\vec{B}/B^2$. The smallness of the Debye length \cite{Boozer:NF3D} implies $\vec{\nabla}\cdot \vec{j}=0$, which is equivalent to
\begin{eqnarray}
&& \vec{B}\cdot \vec{\nabla}\frac{j_{||}}{B}=\vec{B}\cdot\vec{\nabla}\times\frac{\vec{f}_L}{B^2}, \label{j/B force}  \mbox{    where   }\\
&& \vec{f}_L = \vec{j}\times\vec{B} \label{f_L exp}
\end{eqnarray} 
is the Lorentz force.   In a near-ideal plasma, the force implied by Equation (\ref{j/B force}) relaxes by Alfv\'en waves, Appendix \ref{sec:Alfven}.  This process is complicated by the rapid transfer of the energy in the Alfv\'en waves to the plasma through phase mixing in regions in which magnetic field lines exponentiate apart \cite{Heyvaerts-Priest:1983,Similon:1989}.  This exponentiation in the separation is the cause of the exponential enhancement of sensitivity that leads to fast magnetic reconnection.  Alfv\'en-wave damping heats ions through viscosity and electrons through resistivity.  The damping presumably slows the relaxation of $j_{||}/B$, but this has not been studied.  

An Alfv\'enic flattening of $j_{||}/B$ appears consistent with tokamak experiments, but remarkably little has been published on the time empirically required for a drop in the internal inductance $\ell_i$ during tokamak disruptions.  In a toroidal plasma, the flattening of $j_{||}/B$ over the volume covered by a magnetic field line requires a shear Alfv\'en wave  propagate for much greater distance than just the radius of the plasma.  A hundred toroidal transits may be required.

When $j_{||}/B$ is flattened in a toroidal plasma, helicity dissipation, which is due to $\int\mathcal{E}_{ni}Bd^3x$, can become appreciable.  This enhancement of the helicity dissipation is due to two effects: (1) the overall cooling of the plasma and even more importantly (2) the spreading of $j_{||}/B$ into the regions of high resistivity at the plasma edge.  These effects can be modeled using an evolution equation \cite{Boozer:e-runaway2019} for the net plasma current $I(\psi_t,t)$ enclosed in a region containing toroidal magnetic flux $\psi_t$ with $\partial I/\partial\psi_t=j_{||}/B$,
 \begin{eqnarray}
\frac{\partial L I}{\partial t} &=& -2\mathcal{D}[I]; \label{I ev}\\
\mathcal{D}[I]&\equiv& - \psi_t \frac{\partial}{\partial\psi_t}\left\{\mathcal{R}_\psi \frac{dI}{d\psi_t} - \frac{\partial}{\partial\psi_t} \left(\psi_t\Lambda_m\frac{\partial^2 I}{\partial\psi_t^2} \right)\right\}. \hspace{0.1in} \label{Dissipation operator}  \nonumber\\
 \end{eqnarray}
 $L$ is an inductance, $\mathcal{R}_\psi$ is the plasma resistivity, and $\Lambda_m$ models the spreading of the plasma current in regions of stochastic magnetic field lines.  Reference \cite{Boozer:e-runaway2019} gives examples of solutions, which show enhanced resistive decay of  magnetic helicity when the magnetic field is stochastic.


\section{Summary \label{Summary} }

An exponentially increasing sensitivity with time is a mathematical property of the ideal evolution equation  $\partial\vec{B}/\partial t=\vec{\nabla}(\vec{u}_\bot\times\vec{B})$ in three dimensions.  Since this is a mathematical property, it can only be refuted if a fundamental flaw can be found in the mathematics \cite{Boozer:Prevalent2018,Boozer:ideal-ev}.  The same sensitivity is not present in two-dimensional systems, which probably accounts for the effect being overlooked in the reconnection literature.   

An understanding of fast magnetic reconnection requires answers to the four questions that are listed in the Introduction.  This paper addresses the transfer of energy from the magnetic field to the plasma in the near ideal limit of $\mathcal{E}_{ni}\rightarrow0$ and the limitation that helicity conservation places on that transfer.  

Energy can be transferred from the magnetic field to the plasma even in an ideal evolution, but the constraint of an ideal evolution, Equation (\ref{Ideal-ev}), limits the energy that can be transferred.  In a fast magnetic reconnection the energy that can be transferred is limited by magnetic helicity conservation, Section \ref{sec:A.B cons}, which is a weaker constraint.  Helicity conservation implies fast magnetic reconnection process is quasi-ideal in the sense that magnetic helicity can only decay by non-ideal effects such as plasma resistivity.  Early in the post thermal-quench period of tokamak disruptions, a relatively rapid helicity decay occurs due to the breakup of the magnetic surfaces causing both a rapid plasma cooling and a spread of the plasma current into the highly resistive regions at the plasma edge and near surrounding walls \cite{Boozer:e-runaway2019}.

Energy transfer from the magnetic field to the plasma in the limit as $\mathcal{E}_{ni}\rightarrow0$ is partly produced by the non-dissipative transfer of magnetic field energy to Alfv\'en waves, which are then damped on the plasma.  Appendix \ref{sec:Alfven} derives the transfer equation.   Two effects increase the kinetic energy of the motion of particles along the magnetic field $K_{||}=\frac{m}{2}v_{||}^2$.   The more important effect is apparently the exponentiation of $K_{||}$ through the coefficient $\nu_K$, Equation (\ref{nu_K def}),  which can change sign but does not average to zero.  Another is an effective parallel electric field $\mathcal{E}_{||}$, Equation (\ref{eff.E-field}), which is far larger than the true parallel electric field $E_{||}$ when $\mathcal{E}_{ni}\rightarrow0$.   An important term in $\mathcal{E}_{||}$ is the time derivative of the magnetic field that arises from non-ideal effects, $\vec{B}_{ni}$.  This field is derived in \cite{Boozer:ideal-ev} and is shown to be an exponential function of time multiplying a term proportional to $\mathcal{E}_{ni}$ until  $\vec{B}_{ni}$ contributes significantly to the total magnetic field.

The properties of a near-ideal evolution in systems that depend on all three spatial coordinates are fundamentally different from systems that depend on only two as in standard plasmoid theories.   Two-dimensional plasmoid theory dominates modern reconnection studies \cite{Zweibel:review,Loureiro:2016}.  A search on the Web of Science for ``plasmoid magnetic reconnection" returned appoximately five-hundred results.   Nevertheless, the applicability of plasmoid theory to three-dimensional reconnection problems faces a number of mathematical challenges \cite{plasmoid challenges}.

Plasmoid theories do not address the first two questions that are important for understanding fast magnetic reconnection.  (1) Why does a magnetic evolution commonly take an arbitrary initial state into a state of fast magnetic reconnection?  (2) How long does this evolution take?  Plasmoid reconnection theories, \cite{Zweibel:review,Loureiro:2016}, are initiated by a highly unstable narrow current sheet.  Loureiro and Uzdensky recognized in their review of magnetic reconnection by plasmoids that ``the formation of a current sheet and the subsequent reconnection process cannot be decoupled, as is commonly assumed" \cite{Loureiro:2016}.  The second question presupposes an answer to the first.  When 
$\mathcal{E}_{ni}\rightarrow0$, the answer in three-dimensional space is the time to reach a state of rapid reconnection is given by Equation (\ref{tau_trig}).

The language of plasmoids is commonly used to describe reconnection phenomena.  Gou et al \cite{Gou:2019} used plasmoids to describe what they interpreted as regions of strong currents on a scale of a thousand kilometers in their observations of an eruptive region forty-thousand kilometers across in the solar corona.  Transfering their distance scales into kilometers requires knowing that a solar arcsecond is 725~km.  Thousand-kilometer structures are resolved but are approaching the resolution limit.   As noted in the Introduction when exponentiation effects are included, the spatial scale of the primary current sheets is a factor $\ln(R_m)$ smaller than the eruptive region.  In the corona $R_m$ is approximately $10^{12} \approx e^{28}$, so the current sheets observed by Gou et al are on the expected scale.   The natural scales for plasmoids are set by the distance magnetic lines can diffusive resistively, approximately 30 m during the 720 s time scale of the event, or by kinetic scales: the ion gyroradius $\sim20$~cm. the ion skin depth $\sim3$~m, and the electron skin depth $\sim 7$~cm.  Gou et al used plasmoid coalescence to explain the thousand-kilometer scale, but it is far from obvious from models of plasmoid distributions \cite{Huang:2012} that plasmoids would be present on such a large scale. 

The current density in plasmoid theories can exceed the level required for a Dreicer electron runaway \cite{Ji:2011,Cassak:2013}.  Based on the work of Kulsrud et al \cite{Kulsrud:1973}, the rate of runaway becomes important at an electric field more than an order of magnitude smaller than assumed in these papers.  Electron runaway can explain the solar corona p. 1092 of  \cite{Boozer:RMP} and Appendix \ref{corona}.  Standard plasmoid theory may predict a runaway, but plasmoids are not required.  Even simple motions at the footpoints of the magnetic field lines that exit the sun can produce strong localized currents Appendix B of \cite{Boozer:ideal-ev} and have a current density that exceeds the requirements for runaway.

An understanding of structures, such as those observed by Gou et al, would be furthered by numerical simulations, but these simulations must be based on simplified models such as the model given in \cite{Boozer:Prevalent2018}.  As noted in that reference, a realistic simulation of fast magnetic reconnection when $R_m=10^{12}$ is far beyond what is achievable with existing computers.   

The exponentially increasing separation between magnetic field lines is a well-known effect.  It is difficult to understand how excluding this effect, as in standard plasmoid theory, can enhance the reliability of calculations. 


\section*{Acknowledgements}

This material is based upon work supported by the U.S. Department of Energy, Office of Science, Office of Fusion Energy Sciences under Award Numbers DE-FG02-95ER54333, DE-FG02-03ER54696, DE-SC0018424, and DE-SC0019479.



\appendix



\section{Lagrangian coordinates \label{sec:L-coord} }

Lagrangian coordinates and their exponentiation properties are not commonly used in plasma physics but are key concepts in other areas of classical physics in which flows occur.  Two examples are the theory of mixing, which is important from the mixing of food in cooking to the mixing of paints and chemicals, \cite{Aref:2017} and oceanography \cite{Haller:2015}.

Lagrangian coordinates $\vec{x}_0$ are defined so that the position vector in ordinary Cartesian coordinates is $\vec{x}(\vec{x}_0,t)$, where
\begin{equation}
\left(\frac{\partial \vec{x}}{\partial t}\right)_L  \equiv \vec{u}_\bot(\vec{x},t)   \mbox{    with   } \vec{x}(\vec{x}_0,t=0)=\vec{x}_0.
\end{equation}
The subscript ``$L$" implies the Lagrangian coordinates $\vec{x}_0$ are held fixed, $(\partial f/\partial t)_L\equiv(\partial f/\partial t)_{\vec{x}_0}$.


\subsection{Jacobian matrix of Lagrangian coordinates}

The three-by-three Jacobian matrix of Lagrangian coordinates can be decomposed as
\begin{eqnarray}
\frac{\partial \vec{x}}{\partial \vec{x}_0} &\equiv& \left(\begin{array}{ccc}\frac{\partial x}{\partial x_0} & \frac{\partial x}{\partial y_0} & \frac{\partial x}{\partial z_0} \vspace{0.03in}  \\ \vspace{0.03in} \frac{\partial y}{\partial x_0} & \frac{\partial y}{\partial y_0} & \frac{\partial y}{\partial z_0}  \\ \frac{\partial z}{\partial x_0} & \frac{\partial z}{\partial y_0} & \frac{\partial z}{\partial z_0}\end{array}\right) \nonumber\\
  &=& \tensor{U}\cdot\left(\begin{array}{ccc}\Lambda_u & 0 & 0 \\0 & \Lambda_m & 0 \\0 & 0 & \Lambda_s\end{array}\right)\cdot\tensor{V}^\dag . \label{SVD.of.Jacobian}
\end{eqnarray}
where $\tensor{U}$ and $\tensor{V}$ are unitary matrices, $\tensor{U}\cdot\tensor{U}^\dag=\tensor{1}$.   The three real coefficients $\Lambda_u \geq \Lambda_m \geq \Lambda_s \geq 0$ are the singular values of a Singular Value Decomposition (SVD).  The Jacobian matrix can also be written as \cite{Boozer:ideal-ev}
\begin{equation}
\frac{\partial\vec{x}}{\partial \vec{x}_0} = \hat{U}\Lambda_u \hat{u} + \hat{M}\Lambda_m \hat{m} + \hat{S}\Lambda_s \hat{s},
\end{equation}
where $\hat{U}$, $\hat{M}$, and $\hat{S}$ are orthogonal unit vectors, $\hat{U}=\hat{M}\times\hat{S}$, of the unitary matrix $\tensor{U}$, which means they define directions in the ordinary space of Cartesian coordinates.  The unit vectors $\hat{u}$, $\hat{m}$, and $\hat{s}$ are determined by the unitary matrix $\tensor{V}$, which means that they define directions in the space of Lagrangian coordinates.

The Jacobian of Lagrangian coordinates, which is the determinant of the Jacobian matrix, is 
\begin{eqnarray}
J_L&=&\Lambda_u \Lambda_m \Lambda_s; \\
\left(\frac{\partial \ln(J_L)}{\partial t}\right)_L &=& \vec{\nabla}\cdot\vec{u}_\bot,  \label{J-L}
\end{eqnarray}
for a proof see \cite{Boozer:Prevalent2018}.

For almost all velocity fields $\vec{u}_{\bot}$, the coefficient $\Lambda_u$ becomes exponentially large as time progresses,  $\Lambda_s$ becomes exponentially small, and $\Lambda_m$ generally changes slowly with time.  When this is not true, as in an evolution with a symmetry direction, a small perturbation will result in a system in which the singular values have these properties.  In an ideal reduced MHD model, The distribution of parallel current $\mathcal{K}\equiv \mu_0 j_{||}/B$, which is proportional \cite{Boozer:ideal-ev} to $\Lambda_m^2$, is itself proportional to time.   


\subsubsection{Form of $\vec{B}_I$ in Lagrangian coordinates}

An ideal magnetic field $\vec{B}_I$ obeys the evolution equation $(\partial \vec{B}_I/\partial t)_{\vec{x}}=\vec{\nabla}\times(\vec{u}_\bot\times\vec{B}_I)$.  This evolution is represented by the remarkably simple equation, whose history was reviewed by Stern \cite{Stern:1966},
\begin{eqnarray}
\vec{B}_I(\vec{x},t) &=& \frac{1}{J_L} \frac{\partial\vec{x}}{\partial \vec{x}_0} \cdot \vec{B}_0(\vec{x}_0) \\
&=& \frac{\hat{u}\cdot\vec{B}_0}{\Lambda_m\Lambda_s} \vec{U} +  \frac{\hat{m}\cdot\vec{B}_0}{\Lambda_u\Lambda_s} \vec{M} +  \frac{\hat{s}\cdot\vec{B}_0}{\Lambda_u\Lambda_m} \vec{S}.  \hspace{0.2in}  \label{Exp-B}
\end{eqnarray}
The second form, Equation (\ref{Exp-B}), was given in \cite{Boozer:ideal-ev}.  The $t=0$ magnetic field is $\vec{B}_0(\vec{x}_0)$.

A magnetic field that is evolving essentially ideally towards a rapidly reconnecting state generally has a magnitude that neither increases nor decreases exponentially with time.  Consequently, Equation (\ref{Exp-B}) implies the magnetic fields that are of central interest for reconnection studies have the form \cite{Boozer:ideal-ev}
\begin{equation}
J_L \vec{B}_I = \Lambda_m B_0 \hat{M},  \label{B-form}
\end{equation}
so this form will be assumed.  In other words, if the plasma is not exerting an exponentially increasing force on the magnetic field to balance an exponentially increasing magnetic field strength, the velocity of the magnetic field lines must take a form in which $\hat{u}\cdot\vec{B}_0\rightarrow0$.


\subsubsection{Evolution of $\hat{M}$  \label{M ev} }

The ideal evolution of a magnetic field in Lagrangian coordinates is

\begin{eqnarray}
\left(\frac{\partial \vec{B}_I}{\partial t}\right)_{L} &=&\vec{\nabla}\times(\vec{u}_\bot\times\vec{B}_I) + \vec{u}_\bot\cdot\vec{\nabla}\vec{B}_I \hspace{0.2in} \\
&=& \vec{B}_I\cdot\vec{\nabla}\vec{u}_\bot - \vec{B}_I \vec{\nabla}\cdot \vec{u}_\bot ; \\
\left(\frac{\partial J_L\vec{B}_I}{\partial t}\right)_L &=&J_LB_I\cdot\vec{\nabla}\vec{u}_\bot
\end{eqnarray}
since $\vec{\nabla}\cdot \vec{u}_\bot=(\partial \ln J_L/\partial t)_L$.  Using Equation \ref{B-form}, one has $J_L\vec{B}_I= \Lambda_m B_0 \hat{M}$, and
\begin{eqnarray}
\left(\frac{\partial \Lambda_m \hat{M}}{\partial t}\right)_L &=& \Lambda_m \hat{M}\cdot\vec{\nabla} \vec{u}_\bot, \mbox{   so   }\\
\left(\frac{\partial \hat{M}}{\partial t}\right)_{L} &=& (\hat{M}\cdot\vec{\nabla}\vec{u}_\bot)_\bot;  \\
\left(\frac{\partial \ln\Lambda_m}{\partial t}\right)_L &=&\hat{M}\cdot(\hat{M}\cdot\vec{\nabla}\vec{u}_\bot).
\end{eqnarray}.

In an ideal evolution, the unit vector along the magnetic field $\hat{b}_I=\hat{M}$, and
\begin{eqnarray}
\vec{u}_\bot\cdot\left(\frac{\partial \hat{M}}{\partial t}\right)_L&=&\hat{M}\cdot\vec{\nabla} \frac{u_\bot^2}{2},  \label{u.dbI/dt} \\
\end{eqnarray}
In an evolution that differs only by a small amount from an ideal evolution
\begin{equation}
\hat{b} =\hat{M}+\vec{b}_{ni},
\end{equation}
where $\vec{b}_{ni}$ is calculated in \cite{Boozer:ideal-ev}.



\section{$dK/dt$ expression \label{sec:dK/dt}}

\subsection{Guiding-center velocity \label{sec:v_g} }

Northrop's \cite{Northrop:1963} expression for the guiding-center velocity is
\begin{eqnarray}
\vec{v}_g &=& v_{||}\hat{b}+\frac{\hat{b}}{B}\times \Bigg(\vec{\nabla}\Phi - (\vec{E}+\vec{\nabla}\Phi) +\frac{\mu}{q} \vec{\nabla}B \hspace{0.3in}  \nonumber\\
 &&+\frac{m}{q}\frac{d (v_{||}\hat{b}+\vec{u}_\bot)}{dt} \Bigg)  \label{v_g Northrop}
 \end{eqnarray}
Using Equation (\ref{E}) for the electric field, this expression can be written as
\begin{eqnarray}
\vec{v}_g &=& v_{||}\hat{b}+ \vec{u}_\bot + \frac{\hat{b}}{B}\times \Bigg(\vec{\nabla}\Phi - \mathcal{E}_{ni}\vec{\nabla}\ell+\frac{\mu}{q} \vec{\nabla}B \nonumber\\
 &&+\frac{m}{q}\frac{d (v_{||}\hat{b}+\vec{u}_\bot)}{dt} \Bigg). \label{v_g anal}
 \end{eqnarray}
 The total time derivative of the magnetic field direction $\hat{b}\equiv \vec{B}/B$ and the magnetic field line velocity are
 \begin{eqnarray}
 \frac{d\hat{b}}{dt}&=& \left(\frac{\partial \hat{b}}{\partial t}\right)_{\vec{x} }+ (v_{||}\hat{b}+\vec{u}_\bot)\cdot \vec{\nabla} \hat{b} \\
  \frac{d\vec{u}_\bot}{dt}&=& \left(\frac{\partial \vec{u}_\bot}{\partial t}\right)_{\vec{x} }+ (v_{||}\hat{b}+\vec{u}_\bot)\cdot \vec{\nabla} \vec{u}_\bot.
\end{eqnarray}

The total time derivative of the parallel velocity gives not only the curvature drift of particles but also a term along the magnetic field,
\begin{eqnarray}
&&\frac{d(v_{||}\hat{b})}{dt}=\frac{dv_{||}}{dt}\hat{b} + v_{||}^2 (\hat{b}\cdot\vec{\nabla})\hat{b}, \mbox{   where   } \label{dv_||/dt} \\
&&(\hat{b}\cdot\vec{\nabla})\hat{b}\equiv\vec{\kappa},
\end{eqnarray}
which is the curvature of the magnetic field line. 


\subsection{Expression for $dK/dt$}

The electric field enters Northrop's expression for $dK/dt$, Equation (\ref{Northrop dK/dt}), in two ways.  One is implicit, in the expression for $\vec{v}_g$, which was analyzed in Section (\ref{sec:v_g}), and the other is explicit, which will now be considered.

Substituting the electric field from Equation (\ref{E}) into Northrop's expression for $dK/dt$, Equation (\ref{Northrop dK/dt}), one finds
\begin{eqnarray}
\frac{dK}{dt} &=& q\vec{v}_g\cdot \left( -\vec{u}_\bot \times \vec{B} - \vec{\nabla}\Phi + \mathcal{E}_{ni}\vec{\nabla}\ell \right) \nonumber\\
&& +\mu \left(\frac{\partial B}{\partial t}\right)_{\vec{x}}  \label{second dK/dt}
\end{eqnarray}

Using Equation (\ref{v_g anal}) for $\vec{v}_g$ and Equation (\ref{dv_||/dt}) for the time derivative of the particle velocity along the magnetic field, the term in Equation (\ref{second dK/dt})
\begin{eqnarray}
q\vec{u}_\bot\cdot(\vec{v}_g\times\vec{B}) &=&q\vec{u}_\bot\cdot\Bigg(\vec{\nabla}\Phi - \mathcal{E}_{ni}\vec{\nabla}\ell+\frac{\mu}{q} \vec{\nabla}B \nonumber\\
 &&+\frac{m}{q}\frac{d (v_{||}\hat{b}+\vec{u}_\bot)}{dt} \Bigg)_\bot \nonumber \\
 &=& \vec{u}_\bot\cdot\vec{\nabla}(q\Phi +\mu B) +\frac{m}{2} \frac{du_\bot^2}{dt} \nonumber\\
 && +mv_{||}^2 \vec{u}_\bot \cdot\vec{\kappa} - \mathcal{E}_{ni}\vec{u}_\bot\cdot\vec{\nabla}\ell.
 \end{eqnarray}

Equation (\ref{second dK/dt}) for $dK/dt$ can then be written in the form of Equation (\ref{dK/dt eq}).



\section{Expression for $\vec{u}_\bot\cdot\vec{\kappa}$  \label{u.kappa}}

An expression for $\vec{u}\cdot\vec{\kappa}$ can be obtained using the magnetic Poynting theorem.


\subsection{Magnetic Poynting theorem}

Ampere's law plus Faraday's law imply the magnetic Poynting theorem,
\begin{equation}
\left(\frac{\partial}{\partial t}\frac{B^2}{2\mu_0} \right)_{\vec{x}} + \vec{\nabla}\cdot\left(\frac{\vec{E}\times\vec{B}}{\mu_0}\right) + \vec{j}\cdot\vec{E}=0. \label{Poynting eq}
\end{equation}

The electric field is given by Equation (\ref{E}), which can be used to obtain
\begin{eqnarray}
&&\frac{\vec{E}\times\vec{B}}{\mu_0} = \frac{B^2}{\mu_0} \vec{u}_\bot + \frac{\vec{B}}{\mu_0}\times\vec{\nabla}\Phi -\frac{\vec{B}}{\mu_0}\times\mathcal{E}_{ni}\vec{\nabla}\ell;  \hspace{0.2in} \\
&&\vec{\nabla}\cdot\left(\frac{\vec{E}\times\vec{B}}{\mu_0}\right) = \vec{\nabla}\cdot\left(  \frac{B^2}{\mu_0} \vec{u}_\bot  \right) +\vec{j}\cdot\vec{\nabla}\Phi \nonumber\\
&& \hspace{0.3in} - \mathcal{E}_{ni} \vec{j}\cdot\vec{\nabla}\ell + (\vec{B}\times\vec{\nabla}\ell)\cdot\vec{\nabla}\mathcal{E}_{ni}     \label{div(EXB)}
\end{eqnarray}
Inserting Equation (\ref{div(EXB)}) into Equation (\ref{Poynting eq}) and using Equation (\ref{E}) for the electric field gives
\begin{equation}
\left(\frac{\partial}{\partial t}\frac{B^2}{2\mu_0} \right)_{\vec{x}}  + \vec{\nabla}\cdot\left(\frac{B^2}{\mu_0}\vec{u}_\bot\right) + \vec{u}_\bot\cdot\vec{f}_L=(\vec{\nabla}\ell\times\vec{B})\cdot\vec{\nabla}\mathcal{E}_{ni}.
\end{equation}


\subsection{Field line curvature}

The magnetic field-line curvature, $\vec{\kappa} \equiv \hat{b}\cdot\vec{\nabla}\hat{b}$ can be written using $\vec{\nabla}(\vec{B}\cdot\vec{B}/B^2)=0$ as
\begin{eqnarray}
\vec{\kappa} &=& -\hat{b}\times\vec{\nabla}\times\frac{\vec{B}}{B}  \\
&=& \frac{\mu_0}{B^2} \vec{f}_L + \frac{\vec{\nabla}_\bot B}{B}. \label{kappa-f_L}
\end{eqnarray}

Using Equation (\ref{kappa-f_L}) for the Lorentz force the magnetic Poynting theorem can be written 
\begin{eqnarray}
&&\left(\frac{\partial}{\partial t}\frac{B^2}{2\mu_0} \right)_{\vec{x}}  + \vec{\nabla}\cdot\left(\frac{B^2}{\mu_0}\vec{u}_\bot\right) -\vec{u}_\bot\cdot \vec{\nabla}_\bot \frac{B^2}{2\mu_0} = \nonumber\\
&& \hspace{0.6in} - \frac{B^2}{\mu_0}\vec{u}_\bot\cdot\vec{\kappa} +(\vec{\nabla}\ell\times\vec{B})\cdot\vec{\nabla}\mathcal{E}_{ni}. 
\end{eqnarray}
Since
\begin{eqnarray}
&&\vec{\nabla}\cdot\left(\frac{B^2}{\mu_0}\vec{u}_\bot\right) -\vec{u}_\bot\cdot \vec{\nabla}_\bot \frac{B^2}{2\mu_0} \nonumber\\
&& \hspace{0.2in} = \vec{u}_\bot\cdot \vec{\nabla}_\bot \frac{B^2}{2\mu_0} +\frac{B^2}{\mu_0}\vec{\nabla}\cdot\vec{u}_\bot,
\end{eqnarray}
The curvature term can then be written as
\begin{eqnarray}
\vec{u}_\bot\cdot\vec{\kappa} &= & -\frac{\mu_0}{B^2}\Bigg(  \left(\frac{\partial}{\partial t}\frac{B^2}{2\mu_0} \right)_L -(\vec{\nabla}\ell\times\vec{B})\cdot\vec{\nabla}\mathcal{E}_{ni}\Bigg) \nonumber \\
&& - \vec{\nabla}\cdot\vec{u}_\bot . \hspace{0.3in} 
\end{eqnarray}

The divergence of the magnetic field line velocity is related to the Jacobian of Lagrangian coordinates, $\vec{\nabla}\cdot\vec{u}_\bot=(\partial \ln(J_L)/\partial t)_L$, Equation (\ref{J-L}).  Therefore,
\begin{equation}
\vec{u}_\bot\cdot\vec{\kappa} =-\left(\frac{\partial \ln(J_LB)}{\partial t} \right)_L +\frac{ ( \vec{\nabla}\ell\times\vec{B})\cdot\vec{\nabla}\mathcal{E}_{ni} }{B^2/\mu_0}. \hspace{0.3in}
\end{equation}

Equation (\ref{B-form}) for $\vec{B}$ in Lagrangian coordinates implies $\Lambda_m\rightarrow J_LB/B_0$ as $\Lambda_u/\Lambda_s\rightarrow\infty$, where $B_0$ is the magnetic field strength in Lagrangian coordinates at $t=0$.  In the limit $\Lambda_u/\Lambda_s\rightarrow\infty$, but with the magnetic field strength remaining finite,
\begin{equation}
\vec{u}_\bot\cdot\vec{\kappa} =-\left(\frac{\partial \ln(\Lambda_m)}{\partial t} \right)_L +\frac{ ( \vec{\nabla}\ell\times\vec{B})\cdot\vec{\nabla}\mathcal{E}_{ni} }{B^2/\mu_0}. \hspace{0.3in} \label{u-kappa}
\end{equation}



\section{Alfv\'enic relaxation \label{sec:Alfven}}

A fast magnetic reconnection naturally joins lines that have different values of $j_{||}/B$.  When this occurs, a large and sudden variation in $j_{||}/B$ is produced along field lines.  The relation between $\vec{B}\cdot\vec{\nabla}(j_{||}/B)$ and the Lorentz force, the force the magnetic field exerts on the plasma, is given by Equation (\ref{j/B force}).  When the Lorentz force suddenly becomes large, it must be balanced by plasma inertia, $\rho_0\partial \vec{v}/\partial t=\vec{f}_L$, where $\rho_0$ is the plasma density.

Under the assumption that the spatial scale defined by the gradient of $j_{||}/B$ across the magnetic field lines is much shorter than that defined by the gradient of the background properties, it will be shown that Alfv\'en waves are driven and damped.

\subsection{Force equation}

Force balance couples the time derivative of the parallel component of the vorticity of the plasma flow, 
\begin{equation}\Omega=\hat{b}\cdot\vec{\nabla}\times\vec{v},
\end{equation} 
to the variation in $j_{||}/B$ with distance $\ell$ along a magnetic field line,
\begin{equation}
 \frac{\partial\Omega}{\partial t} \approx V_A^2 \frac{\partial}{\partial\ell} \frac{\mu_0j_{||}}{B} +\nu_v \nabla^2\Omega; \label{d omega/dt}
 \end{equation}
 $V_A^2 \equiv B^2/\mu_0\rho_0$ is the Alfv\'en speed and $\nu_v$ is the kinematic viscosity of the plasma.
 
 The derivation of the force-balance equation is based on $\rho_0\partial \vec{v}/\partial t=\vec{f}_L$, where $\vec{f}_L=\vec{j}\times\vec{B}$ is the Lorentz force.  The Lorentz force and $\vec{B}\cdot\vec{\nabla}(j_{||}/B)$ are related by $\vec{\nabla}\cdot\vec{j}=0$, which implies, Equation (\ref{j/B force}),
\begin{eqnarray}
\vec{B}\cdot \vec{\nabla}\frac{j_{||}}{B}&=& \vec{B}\cdot\vec{\nabla}\times\frac{\vec{f}_L}{B^2} \\
 &=& \vec{B}\cdot \vec{\nabla}\times\left( \frac{\rho_0 \partial \vec{v}/\partial t}{B^2}-\rho_0\nu_v \nabla^2\vec{v}\right) \hspace{0.2in}\\
&\approx& \frac{\rho_0}{B}  \frac{\partial\Omega}{\partial t} -\rho_0\nu_v \nabla^2\Omega
\end{eqnarray}
Note $\vec{B}\cdot\vec{\nabla}=B\partial/\partial\ell$.

\subsection{Parallel current equation}

Faraday's law and Ohm's law, $\vec{E}+\vec{v}\times\vec{B}=\eta \vec{j}_{||}$, imply that
\begin{equation}
\frac{\partial \Omega}{\partial\ell}\approx \frac{\partial}{\partial t} \frac{\mu_0j_{||}}{B} -\eta \nabla^2\frac{j_{||}}{B}. \label{ d Omega/dz}
\end{equation} \vspace{0.1in}

When $\eta=0$, the velocity of the magnetic field lines is also $\vec{v}$.  The angle $\theta$ that a perturbed magnetic field line winds around an unperturbed line is given in cylindrical coordinates by $v_\theta=r \partial\theta(\ell,t)/\partial t$, so $\Omega = 2 \partial\theta/\partial t$ and $\theta=\mu_0j_{||}/2B$ 

The derivation of Equation (\ref{ d Omega/dz}) is
\begin{eqnarray}
\vec{B} &=& B \hat{b} + \vec{\nabla}\times (A_{||}\hat{b}); \\
\mu_0 j_{||} &\approx& -\nabla_\bot^2 A_{||} \mbox{  from  } \vec{\nabla}\times\vec{B}=\mu_0 \vec{j};  \label{d^2A}\\
\vec{E} &=& - \frac{\partial A_{||}}{\partial t}\hat{b} - \vec{\nabla}\Phi;\\
&=& - \vec{v}\times \vec{B} +\eta\vec{j}_{||}, \mbox{   so  } \\
\frac{\partial A_{||}}{\partial t} &=& -\hat{b} \cdot\vec{\nabla}\Phi -\eta j_{||}\mbox{   and   }\label{ dA/dt}\\
\vec{v} &=& \frac{\vec{B}\times\vec{\nabla}\Phi}{B^2};\\
\Omega &\approx& \frac{\nabla_\bot^2\Phi}{B} \label{d^sPhi}
\end{eqnarray}
Equations (\ref{d^2A}), (\ref{ dA/dt}), and (\ref{d^sPhi}) imply Equation (\ref{ d Omega/dz}).

\subsection{Alfv\'en waves}

The $\ell$ derivative of Equation (\ref{d omega/dt}) and the $t$ derivative of Equation (\ref{ d Omega/dz}) give the equation for an Alfv\'en wave,
 \begin{eqnarray}
&&\frac{\partial}{\partial\ell}\left(V_A^2\frac{\partial}{\partial \ell}\left(\frac{j_{||}}{B}\right)\right) -  \frac{\partial^2}{\partial t^2}\left(\frac{j_{||}}{B}\right)=  \nonumber\\
&& \hspace{0.2in}-\left(\nu_v+\frac{\eta}{\mu_0}\right)\nabla_\bot^2 \frac{\partial}{\partial t}\left(\frac{j_{||}}{B}\right)+\nu_v\frac{\eta}{\mu_0} \nabla_\bot^4\frac{j_{||}}{B}.  \hspace{0.4in}\label{Driven Alfven}
 \end{eqnarray}
 When magnetic field lines exponentiate apart, the Laplacian $\nabla_\bot^2$ becomes exponentially large. 
 
 
 
 \section{Runaway currents \label{corona}}
 
 The solar corona has an obvious explanation if the current density in the transition region of the sun exceeds the Dreicer current, the current required for electron runaway, p. 1092 of  \cite{Boozer:RMP}.  If the Dreicer current is exceeded, electrons runaway to whatever energy is required to carry the current, which means at a sufficiently high energy that the electron density $n$ does not become too small due to the gravitational acceleration of the sun $g$.  When the temperature $T$ is constant  $dn/dr=-n/h$, where the scale height $h\equiv T/Mg$.  When the ionization is high, $M=m_i$, the proton mass, and $h\approx 350T$~km/eV.  A coronal temperature of 100~eV is consistent with a scale height of 35,000 km.  
 
 Song \cite{Song:2017} gives the electron temperature, 1~eV, and the electron number density, $n=3\times 10^{10}/$m$^3$, at the solar transition. These parameters imply the electron thermal speed is $v_e\approx4\times10^5$~m/s and the rate of resistive diffusion is $\eta/\mu_0 \approx 10^3~$m$^2/$s.  When the magnetic field lines are assumed to have a speed of $u_\bot=10^2$~m/s, which would move the lines across a 40,000 ~km wide feature in five days, the width of the region relaxed by resistivity is $\delta_\eta =\eta/\mu_0u_\bot\approx 10$~m. 
 
 The calculations of Kulsrud et al \cite{Kulsrud:1973} imply the rate of electron runaway reaches a significant value when the current density $j_D=2\times10^{-2} e n v_e \approx 38 $~A/m$^2$.  The change in the magnetic field produced by the current density $j_D$ acting over the distance $\delta_\eta$ is $\delta B =\mu_0 J_D \delta_\eta \approx 5$~G, which is of order 1\% of the magnetic field that is present.
 
 When the current density equals the Dreicer current, the rate of energy dissipation per unit volume is large, $\eta j_D^2 \approx 1.4~$W/m$^3$.   The energy density of a magnetic field of $B=5\times10^{-2}$~T is $B^2/2\mu_0\approx10^3$J/m$^3$.  The resistive dissipation could dissipate the magnetic field in 625 s, but the power is supplied by whatever forces drive the complex field line motions a few thousand kilometers below the transition region.  These forces relax along the magnetic field lines Alfv\'enically, Appendix \ref{sec:Alfven}, which is rapid compared to 625 s.  As noted in  \cite{Boozer:RMP} any star that has evolving magnetic field structures on the scale of tens of thousands of kilometers must have a corona, otherwise the induced currents would run out of current carriers, but whether this can give the correct height of the transition region is not clear.  Simulations could clarify the question.
 

\end{document}